\documentclass{aa}  
\usepackage{graphicx}
\usepackage{natbib}
\bibpunct{(}{)}{;}{a}{}{,} %
\usepackage{txfonts}
\newcommand{\igr}{IGR\,J00234+6141}

\newcommand{\ecsa}{\mbox{$\rm ergs\;cm^{-2}s^{-1}\mbox{\AA}^{-1}$}}
\newcommand{\ecs}{\mbox{$\rm ergs\;cm^{-2}s^{-1}$}}
\newcommand{\es}{\mbox{$\rm ergs\;s^{-1}$}}

\newcommand{\gs}{\mbox{$\rm g\;s^{-1}$}}

\newcommand{\kms}{\mbox{$\rm km\,s^{-1}$}}
\newcommand{\Ha}{H$_{\alpha}{ }$}
\newcommand{\Hb}{H$_{\beta} $ }
\newcommand{\Hg}{H$_{\gamma}{ }$ }

\newcommand{\HeII} {HeII(~4686\,\AA) }

\newcommand{\draft}[1]{
}
\draft{Draft 1.0, \today}
\begin{document}
   \title{IGR J00234+6141 : a new INTEGRAL source identified as an Intermediate polar
\thanks{Based on observations obtained at the Haute-Provence Observatory (France) and
at the Loiano Observatory (Italy) operated by the Istituto Nazionale di Astrofisica.}}

   \titlerunning{IGR J00234+6141 : a new hard X-ray IP}

   \author{J.M. Bonnet-Bidaud\inst{1}
   \and D. de Martino\inst{2}
    \and M. Falanga\inst{1}
   \and M. Mouchet\inst{3}
   \and N. Masetti\inst{4}
   }
       
   \offprints{J.M. Bonnet-Bidaud, email: bonnetbidaud@cea.fr}

   \institute{Service d'Astrophysique, DSM/DAPNIA/SAp CE Saclay, 91191 Gif-sur-Yvette, France\\   
    \email{bonnetbidaud@cea.fr}
\and
    INAF--Osservatorio Astronomico di Capodimonte, Via
    Moiariello 16, I-80131 Napoli, Italy\\
    \email{demartino@na.astro.it}
\and
   APC, UMR 7164, University Denis Diderot, 2 place Jussieu, F-75005 and 
   LUTH, Observatoire de Paris, F-92195 Meudon Cedex, France\\
   \email{martine.mouchet@obspm.fr} 
\and
    INAF--Istituto di Astrofisica Spaziale di Bologna, Via
    Gobetti 101, I-40129 Bologna, Italy\\
    \email{masetti@iasfbo.inaf.it}
 }        

   \date{Received 15 May 2007; accepted 3 June 2007 }

 
  \abstract
   {Following an extensive survey of the galactic plane by the INTEGRAL satellite, new hard X-ray sources are discovered with a significant fraction of Cataclysmic Variables (CVs) among them.}
   {We report here the identification of one of these hard X-ray sources, IGR J00234+6141, as an 
accreting magnetic white dwarf of intermediate polar type. }
   {We analyse the high energy emission of the INTEGRAL source using all available data and provide complementary optical photometric and spectroscopic data obtained respectively in August and October  2006.}
   {Based on a refined INTEGRAL position, we confirm the proposed optical identification. We clearly detect the presence of a 564\,s periodic optical modulation that we identify as the rotation of the white dwarf. The analysis of the optical spectrum also demonstrates that the emission lines show a  modulation in radial velocity with an orbital period of $P_{orb} = (4.033 \pm 0.005)$ hr.}
   {The two periodicities indicate that \igr\, is a magnetic CV of the intermediate polar type. This is one of the faintest and hardest sources of this type detected by INTEGRAL. This confirms earlier conclusions that IPs contribute significantly to the population of galactic X-ray sources and represent a significant fraction of the high energy background.}


   \maketitle
%

\section{Introduction}
The first complete galactic plane scan by the INTEGRAL satellite at energy above 20\,keV has shown that Cataclysmic Variables (CVs) constitute a significant fraction of the galactic high energy emission. 
The 2nd INTEGRAL source catalogue, from a $\sim$10\,Ms dataset \citep{bird06}, lists 8 confirmed CVs out of a total of 209 detected sources, amounting therefore to 4\% of the catalogue. The last more sensitive survey, based on a 40\,Ms exposure time \citep{bird07}, has increased the catalog to 421 sources and more than doubled the  total number of CVs (21), increasing their fraction to  5\%. 
Most of the identified high energy CVs are Intermediate Polars (IPs) i.e. accreting magnetic white dwarfs with intermediate ($10^{6}$-$10^{7}$G) magnetic fields.  
Though the standard accretion column model predicts high energy emission of magnetic CVs (mCVs) \citep{aizu73} , only few sources have been studied above  20\,keV {\citep {matt00, demar01, demar04, falanga05}}.
In the strongly magnetized (B$\geq 10^{7}$G) polar systems, cyclotron cooling is an important mechanism  to suppress the bremsstrahlung high temperature emission whilst it should be negligible among IPs. This could explain why most hard X-ray CVs are IPs. However there is so far no clear explanation of the high temperature achieved in some cases and this remains an open problem until a detailed  characterization of each of these high energy CVs is obtained. Here we report on the identification of one of the very few CVs discovered by its high energy emission. \\
IGR J00234+6141 was  first discovered from the INTEGRAL deep survey 
in the Cas-A region \citep{denhartog06}. 
From a 1.6 Ms exposure, a weak ( $\sim0.7$ mCrab) source was detected in the (20-50 keV) band with
significant flux detected up to 100 keV. The best source position was provided with an estimated 
3\arcmin\, accuracy, marginally consistent with a ROSAT soft X-ray counterpart, 
1RXS J002258.3+614111, suggesting an X-ray binary {\citep {denhartog06}. 
Optical investigation of the much smaller ($\approx$10\arcsec) ROSAT error box led to the identification of a (R $\approx$\,16.5) object showing strong emission lines of the Balmer series typical of cataclysmic variables {\citep {halpern06, bikmaev06, masetti06}}. 
From a preliminary short 2\,hr photometric R-band observation, the presence of a possible 570 s periodic modulation was also reported suggesting a rotating white dwarf \citep{bikmaev06}.
We present here the first detailed photometric and spectroscopic observations leading to the secure identification of the source as an Intermediate polar.
%
\section{INTEGRAL analysis}
The INTEGRAL \citep{winkler03} high energy emission (20--100 keV) of IGR~J00234+6141 was analyzed using all publicly-available IBIS/ISGRI \citep{u03,lebr03} data from March 2003 to October 2006. The data were collected for all ISGRI pointings within $\leq 9^{\circ}$ for a total effective exposure $\sim 3.6$~Ms. To study the weak persistent X-ray emission, the ISGRI (20--100 keV) spectrum has been extracted from the mosaic images in four energy bands (20-28 keV), (28-40 keV), (40-57 keV) and (57-80 keV). The data reduction was performed using the standard Offline Science Analysis (OSA) software version 5.1. 
\subsection{Source position}
The source is clearly detected in all energy bands with a maximum significance of 6.8 $\sigma$ in the (28-40 keV) band and a total (20-80 keV) significance of 8.9 $\sigma$. Based on the total exposure of $\sim 3.6$~Ms, the best position is determined to be $\alpha = 00^{h} 22^{m} 54.3{s},\,  \delta = +61\degr 43\arcmin 05\arcsec$ (J2000) with a 90\% accuracy of 5.2\arcmin. This is consistent with the position derived in  the third INTEGRAL catalogue from a lower 1.8~Ms exposure \citep{bird07}. The INTEGRAL position is now in better agreement with the proposed optical counterpart than the first determination by \citet{denhartog06}, reducing the separation from 3.2\arcmin\, to 1.3\arcmin\, (see Fig.~\ref{fig_pos}). \\
We also note that contrary to what was reported by \citet[ see their Fig.~6]{bikmaev06} the position of the ROSAT source, 1RXS J002258.3+614111, is fully consistent with the optical counterpart with an error circle much smaller than shown by the authors and an X-optical separation less than 6.1\arcsec. The identification of \igr with both the soft X-ray source and the optical star is therefore quite secure. 
\begin{figure}[htbp]
 \centering
\includegraphics[height=8.5cm,width=8.5cm]{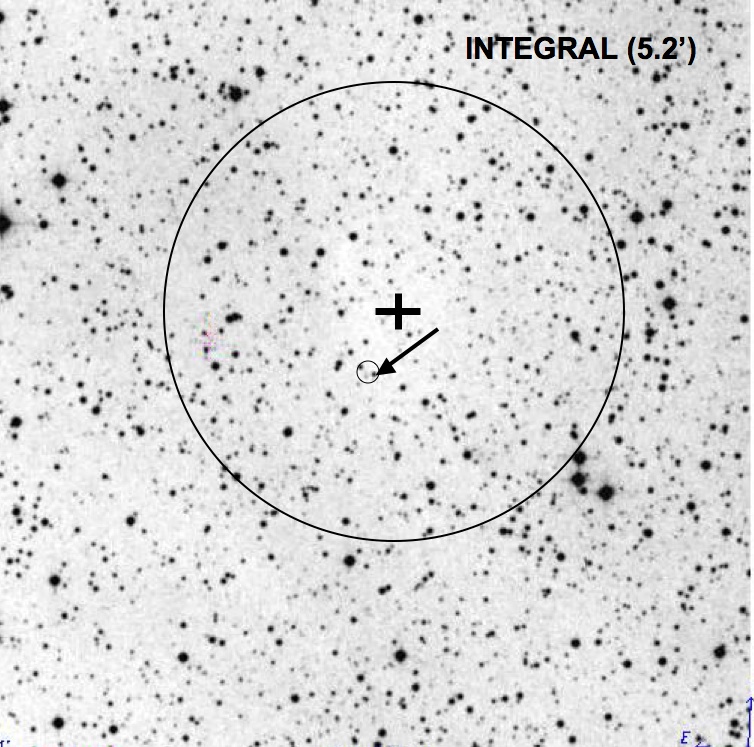}
\caption{New INTEGRAL position and uncertainty for the high energy source \igr, derived from our $\sim 3.6$~Ms exposure (cross and outer 5.2\arcmin\, circle). The inner (11\arcsec\, radius) circle is the ROSAT position, fully consistent with the proposed optical counterpart (arrow).}
\label{fig_pos}
\end{figure}
\begin{figure}
\centering
\includegraphics[width=7.5cm]{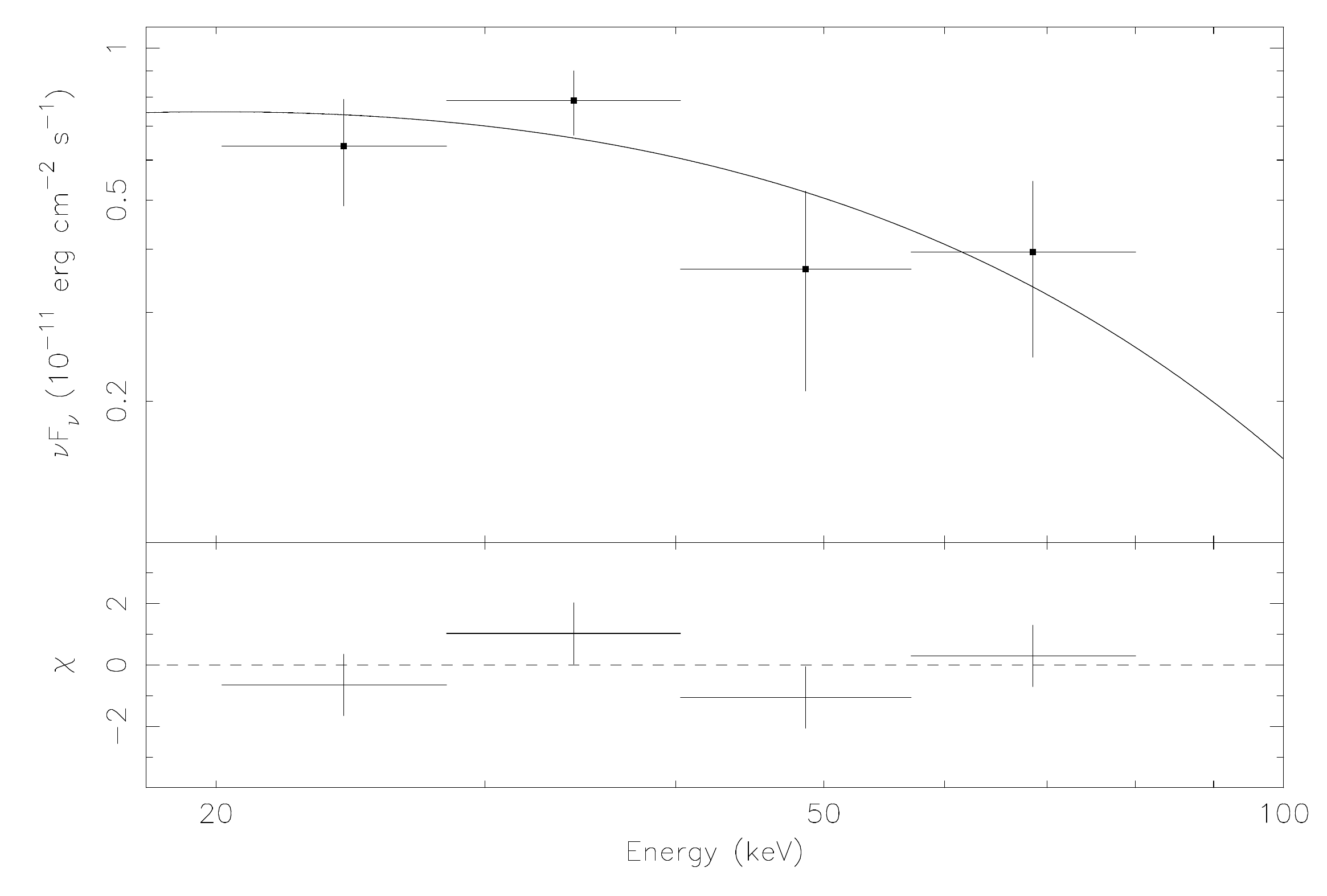}
\caption{The hard $\nu F_{\nu}$  energy spectrum of \igr\, with the best $kT \approx 31$ keV  bremsstrahlung fit superposed and corresponding residuals (bottom).}
\label{fig_spec}
\end{figure}
\subsection{Source spectrum}
Spectral analysis was performed with XSPEC version 12.0 \citep{arnaud96, dorman03}. A systematic error of 2\% was applied to the ISGRI spectrum, which corresponds to the current uncertainties in the response matrix. All uncertainties in the spectral parameters are given at a 90\% confidence level for single parameters. We first fit the spectrum using a simple power-law model which gives a fit with $\chi^{2}{\rm /d.o.f.} = 3.1/2$, and the power-law index is determined to be $\Gamma =2.4^{+0.7}_{-0.6}$. A better fit was found with a bremsstrahlung model more commonly used for CVs, this gave a $\chi^{2}{\rm /d.o.f.} = 2.6/2$, with a bremsstrahlung temperature $kT = 31.3^{+53}_{-14}$ keV. The flux in the (20--100 keV) range is $8.3\times10^{-12}$ erg cm$^{-2}$ s$^{-1}$. In Fig. \ref{fig_spec}, we show the unfolded spectrum and the residuals of the data to the bremsstrahlung model.\\
We find significant differences with the source parameters published from a smaller dataset by \citet{barlow06}, with a factor 2 higher bremsstrahlung temperature and with a factor 1.3 lower flux. But our reduced $\chi^{2}{\rm /d.o.f.} = 2.6/2$ is much more consistent than the one quoted by these authors ($\chi^{2}{\rm /d.o.f.} = 6.1/3$). For the energy band (20-40 keV), we also have a higher significance of $8.2 \sigma$ ($6.4\sigma$ for Barlow et al. 2006) that can be explained since we used only the pointings where the source was within $\leq 9^{\circ}$, therefore increasing the S/N.
\section{Photometric observations}
\igr\,  was observed on August 26 and 27, 2006 
at the  1.5\,m Loiano telescope (Bologna, Italy) equipped with the BFOSC 
detector  operated with the Gunn g filter  \citep{gualandi01}. The integration time was set 
to 30\,s in all observations. The CCD was windowed to a size of 
$\sim5^{\arcmin}$$\times$\,5$^{\arcmin}$ in order to reduce read-out time,
giving a temporal resolution of 40\,s. On August 26, the sky 
conditions were very poor, with winds and clouds and a seeing $\sim$ 
3$^{\arcsec}$ so that the source was observed for  only $\sim$40\,min. On August 27, the 
sky conditions were photometric with a seeing of 1.5-2$^{\arcsec}$, 
although some cirrus were present during the night. We here then present 
the observations obtained on August 27. \\
A total of 455 frames were acquired in 5.3\,hr with the log of the observations 
reported in Table~\ref{obslog}. 
The photometric data were reduced using {\em iraf} package with 
standard procedures including bias, flat-field and sky subtraction. 
Aperture 3$\arcsec$ photometry was obtained for the target and several comparison 
stars. The flux ratios were obtained by dividing the counts of the target 
by the best combination (weighted mean) of three reference stars.   
These ratios were converted to fractional intensities by  dividing by the 
mean flux ratio and heliocentric correction was also applied. 
The resulting light curve, shown in Fig.~\ref{ph_lc} (top panel), displays clear short term periodic variations  superposed on a  longer term variability. 
\begin{table*}[t!]
\caption{Log of photometric and spectroscopic observations.}
\label{obslog}
\centering
\begin{tabular}{c c c c r r }
\hline \hline
\noalign{\smallskip}
Observations &  Range     & Res. &  Date   & UT(start) & Exposure (min)\\
\noalign{\smallskip}
\hline
\noalign{\smallskip}
Photometry   & g filter & 40\,s    & 2006 Aug. 27 & 21:45 & 320  \\
\noalign{\smallskip}
            
Spectroscopy  & 3600-7200 & 5.7\,\AA & 2006 Oct. 21 & 18:56 & 244  \\
                           & 3600-7200 & 5.7\,\AA & 2006 Oct. 24 & 22:34 & 134  \\
                           & 3600-7200 & 5.7\,\AA & 2006 Oct. 25 & 18:24 & 538  \\

\noalign{\smallskip}
\hline
\end{tabular}
\end{table*}
   \begin{figure}
   \centering
   \includegraphics[height=10.0cm,width=8.5cm]{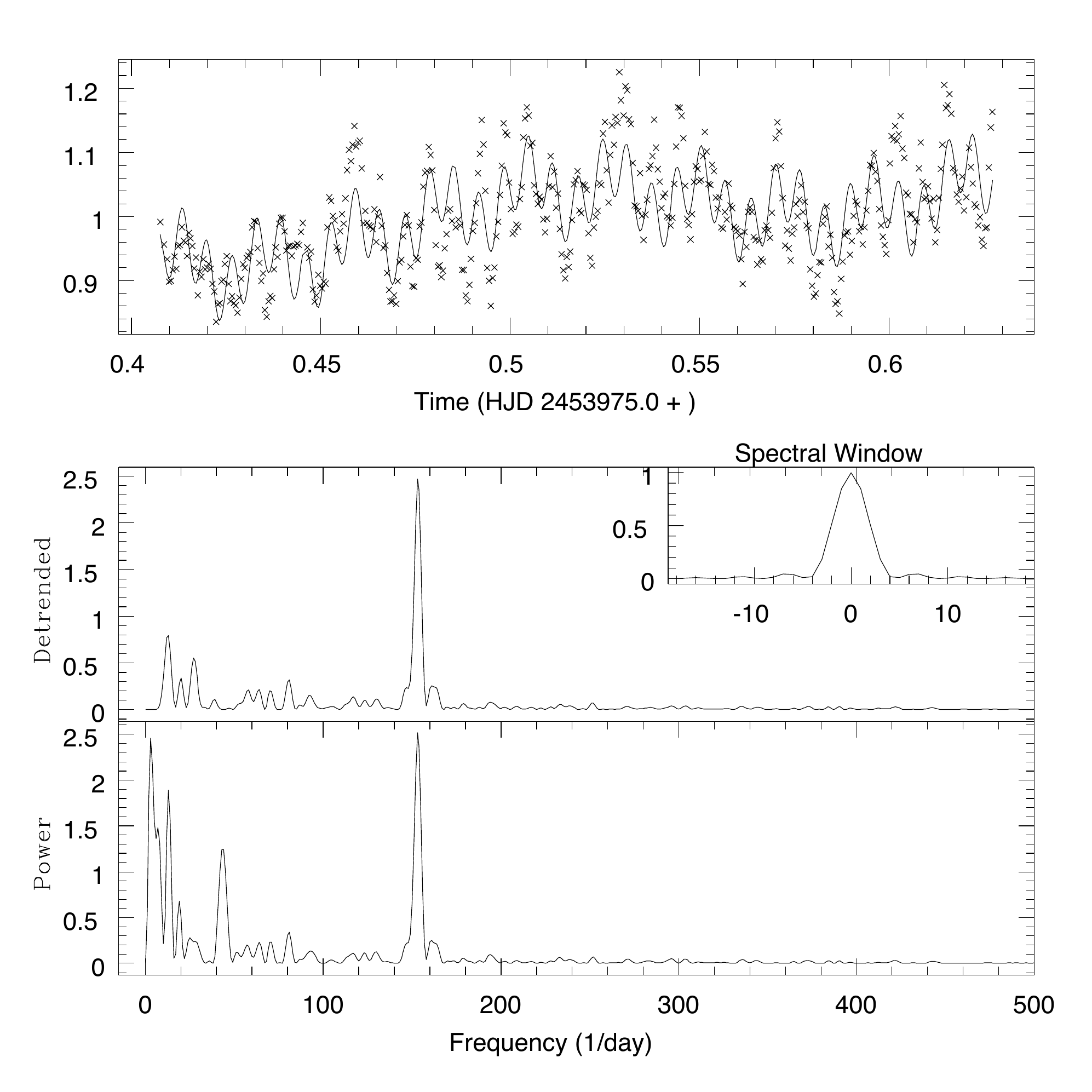}
\caption{Top : \igr\, g-band light curves (points) overlaid with a simulated curve (line)
including the best spin period (564 s) and the long term variations (see text for details). 
Middle : Power spectrum (DFT) of the light curve, detrended from the low-frequency variations.
Ordinate are power amplitudes in units of $10^{-3}$. The inserted panel shows the spectral window
Bottom :  DTF of the observed light curve showing a strong peak at $\sim$ 153\,\,day$^{-1}$. 
Some additional power is also present at lower frequencies (see text).} 
\label{ph_lc}
\end{figure}
A Discrete Fourier Transform
(DFT) on  the photometric time series shows a strong peak at 153\, day$^{-1}$ 
(see Fig.~\ref{ph_lc}, bottom panel).
The low frequency part also shows some additional power  at  $\sim$44\, day$^{-1}$ and 
below 10\,day$^{-1}$. This latter 
could be due to a mixture of residual atmospheric  extinction and  sky transparency 
fluctuations as well as possible contribution from variability at the orbital period 0.168\,day (see below). 
We have simulated the long term variability using a third order polynomial 
function and two sinusoids at the fixed orbital 
frequency and another left free to vary. This latter is then found at 
43.88$\pm$0.23\,day$^{-1}$ (32.82$\pm$0.17\,min).\\
   \begin{figure}
   \centering
   \includegraphics[width=6.0cm]{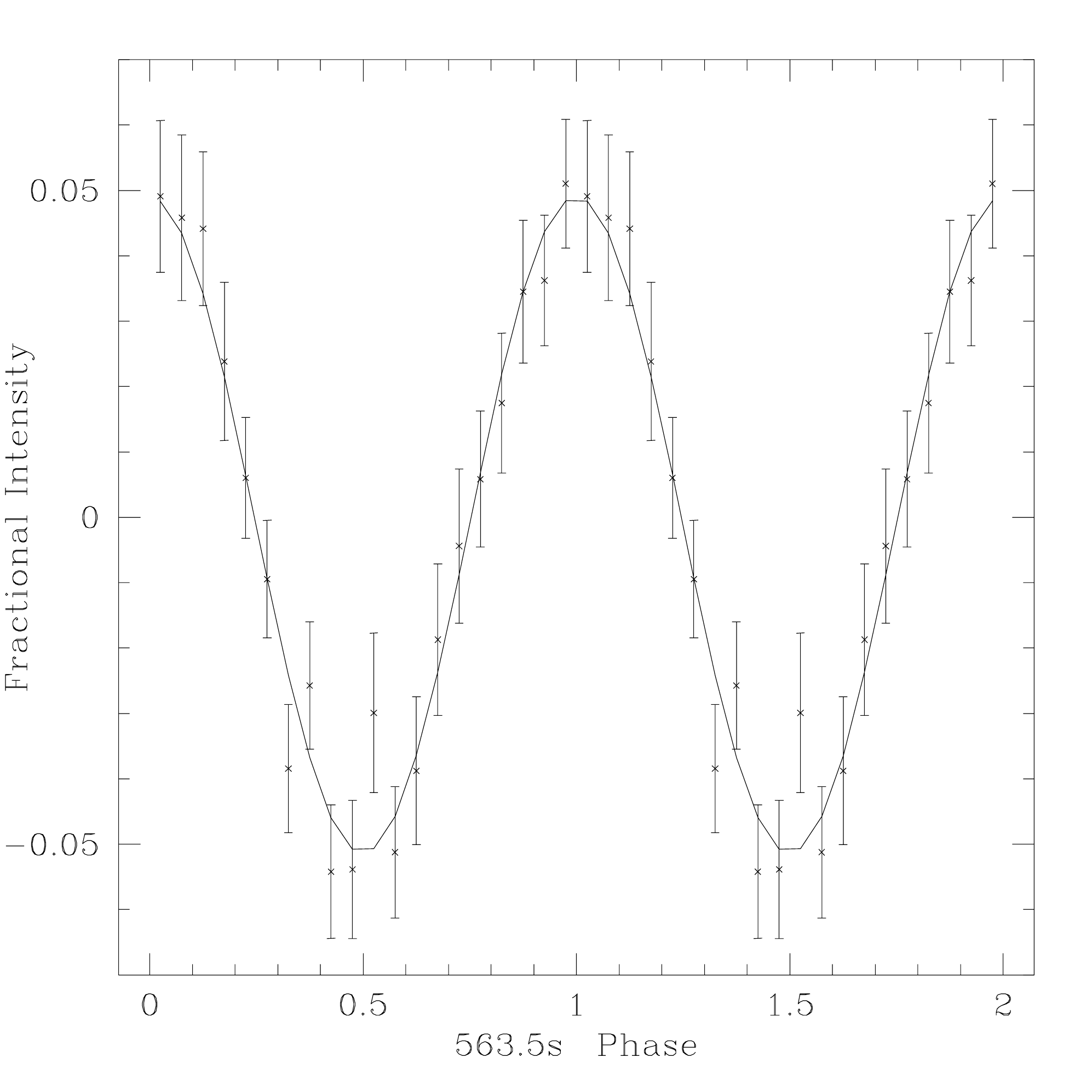}
\caption{The photometric g-band light curve folded with the best 563.5 s ephemeris.
The modulation is very sinusoidal with a full amplitude of 0.10 mag. (full line).} 
\label{ph_fold}
\end{figure}
The DFT of the detrended light curve is shown 
in the middle panel of Fig.~\ref{ph_lc}. 
From a  sinusoidal fit to the detrended curve, the main peak is found at 153.319$\pm$0.167\, day$^{-1}$. 
A periodic 563.53$\pm$0.62\,s  modulation is then clearly and safely detected 
with an ephemeris as:
\begin{equation}
\rm T_{\omega}^{max} = HJD\,2\,453975.51782(7) + 0.006522(7)\,E,
\end{equation}
where $\rm T_{\omega}^{max}$ is the heliocentric arrival time of the sine curve maximum.
There is no indication of beat periods or harmonics.
The light curve folded with this period is indeed very sinusoidal with an amplitude of ($0.100\pm 0.003$) mag (see Fig.~\ref{ph_fold}). 
This periodicity, typical of a spin period, indicates that the system is an intermediate polar. 
Though there appears to be  an indication of a possible 
variability at 32.8\,min,  the identification of this type of longer period variations obviously requires 
a longer coverage.
%
\section{Spectroscopic Observations}
\igr\, was observed at the Observatoire de Haute Provence (France)
in October 2006 (see Table~\ref{obslog}). 
Long slit spectra were obtained with the Carelec spectrograph 
 \citep{lemaitre90}, attached to the Cassegrain focus of the 
193\,cm telescope and using  a CCD EEV (2048x1024 pixels) detector of 
13.5 $\mu$m pixel size.
A 133\,\AA/mm grating was used, with a slit width of 2\arcsec\,,
leading to a  wavelength coverage of 3600-7200\,\AA\, at a FWHM resolution of
$\sim$ 5.7\,\AA. 
Exposure times were set at 1128\,s, twice the 564\,s modulation,
to smear any effect from the spin variability. 
The observations were obtained in nearly photometric conditions 
with only partial thin clouds during one night (21 Oct.). Seeing was typically of 2.5\arcsec\, 
 to 3\arcsec\,.

Standard reduction was performed in the ESO-MIDAS package, including
cosmic rays removal, bias subtraction, flat-field correction and wavelength 
calibration. The wavelength calibration was checked on sky lines which were found within 0.6\,\AA\, from their expected wavelengths. 
All radial velocity measurements have been corrected from the Earth motion and 
from the small instrumental shifts measured on the OI 5577\,\AA\, line
and times have been converted in the heliocentric system.
Flux calibration has been performed using the standards BD+28 4211 and EG 247.

The mean optical spectrum of \igr\, 
 is typical of magnetic CVs with strong emission lines of the Balmer series, HeII (4686\AA),   HeI (4471, 5875, 6678 and 7065\AA) and CIII-NIII (4655\AA), superimposed on a relatively blue continuum. 
The mean V flux is estimated at $\sim 7.5 \times 10^{-16}$ \ecsa , which corresponds to a V magnitude of $\sim$16.7, consistent to that reported by {\citet {bikmaev06}} and {\citet {masetti06}}. 
%
   \begin{figure}
   \centering
\includegraphics[width=6.0cm,angle=-90]{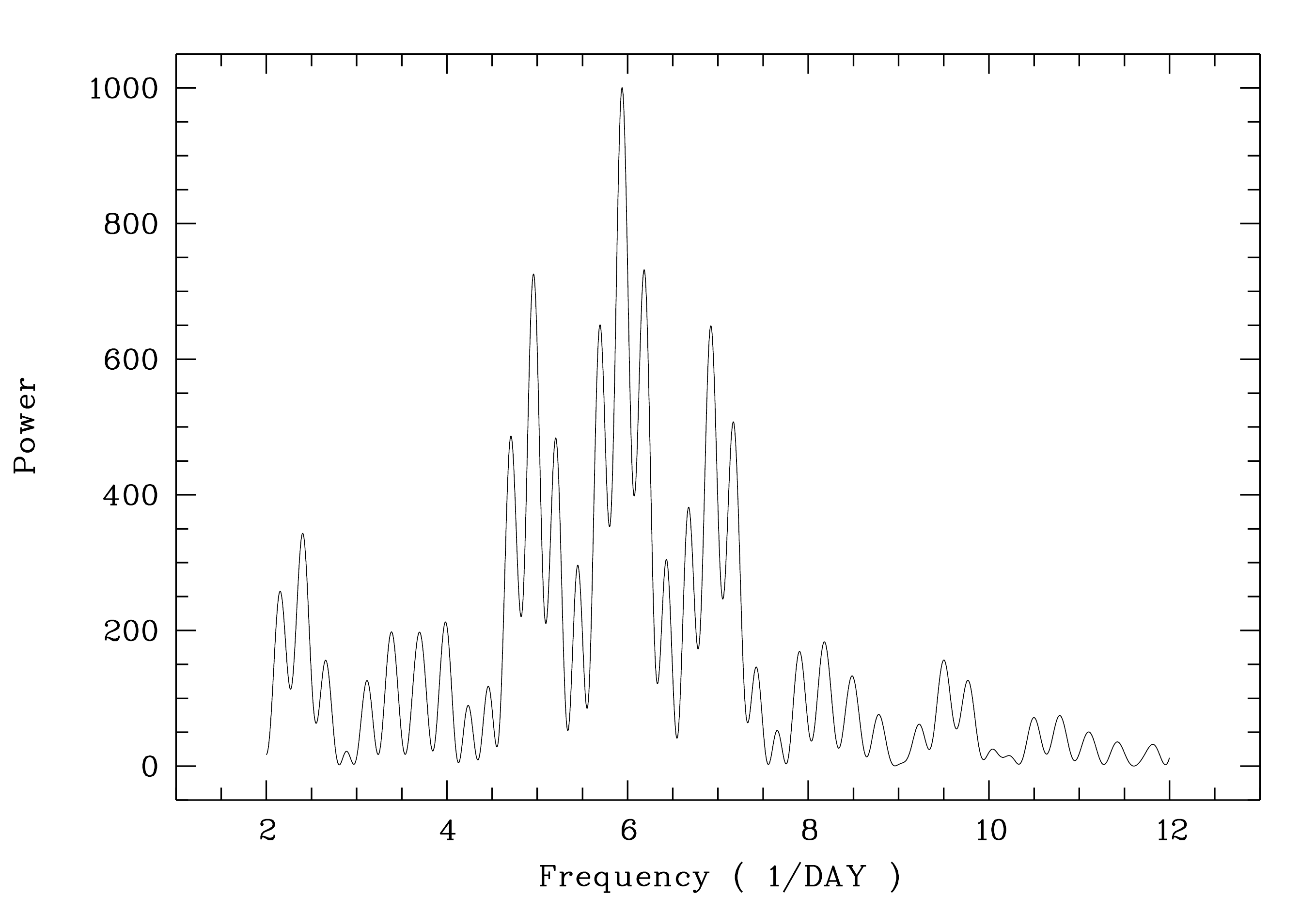}
\caption{Power spectrum of the  \Ha\, radial velocities.
The best value is found at
$\rm P_{\Omega} \sim$  5.9\,\,day$^{-1}$ ($\sim$4.1\,hr). Secondary maxima are 1-day and 4-day harmonics.}
\label{sp_power}
\end{figure}
The radial velocities for the main lines were measured using a single Gaussian least square fit procedure and were analysed to constrain the orbital period.  
Periodicities were searched using a Discrete Fourier Transform (DFT).
The power spectrum for the strongest  \Ha\, line gives the most accurate period determination (see Fig~\ref{sp_power}).  The maximum power is seen at 5.94$\pm$0.14\, day$^{-1}$ (4.03$\pm$0.10\, hr) with secondary peaks which are 1\,d and 4\,d harmonics, corresponding to the night separations. 
The characteristics of the line modulation were further determined from a $\chi^{2}$ sine fit.
The minimum $\chi^{2}$ corresponds to a period of
$\rm P_{\Omega}$=4.033$\pm$0.005\,hr, where the error bar is at a 99\% level, computed for two independent parameters.  
 The orbital ephemeris for \igr\, is determined as : 
\begin{equation}
\rm T_{\Omega}= HJD\,2454033.4717(12) + 0.16804(21)\,E,
\end{equation}
where $\rm T_{\Omega}$ is the predicted heliocentric time of the blue-to-red radial velocity transition.

\begin{table}[t!]
\caption{The parameters of the orbital modulation of the emission lines radial velocities.}
\label{velocity}
\centering
\begin{tabular}{l l l r }
\hline \hline
\noalign{\smallskip}
Lines &  $\gamma$ ( \kms )  & K ( \kms ) &  Phase (*)  \\
\noalign{\smallskip}
\hline
\noalign{\smallskip}
\Hg    & - 24.5 (5.2)    & 47.1 (6.9)    & - 0.07 (0.07)   \\
\HeII  & - 60.1 (8.7)    & 71.3 (11.7)  &  +0.08 (0.09)  \\
\Hb    & - 30.6 (4.5)    & 53.9 (6.1)    & - 0.07 (0.06)   \\
\Ha    & - 31.6 (3.0)    & 60.0.(4.0)    &   0.0   (fixed)   \\           
\noalign{\smallskip}
\hline
\multicolumn{4}{l}{* Phase of the blue-to-red zero crossing }\\
\multicolumn{4}{l}{1-$\sigma$ error bars into parentheses }\\
\end{tabular}
\end{table}
   \begin{figure}
   \centering
\includegraphics[width=8.5cm]{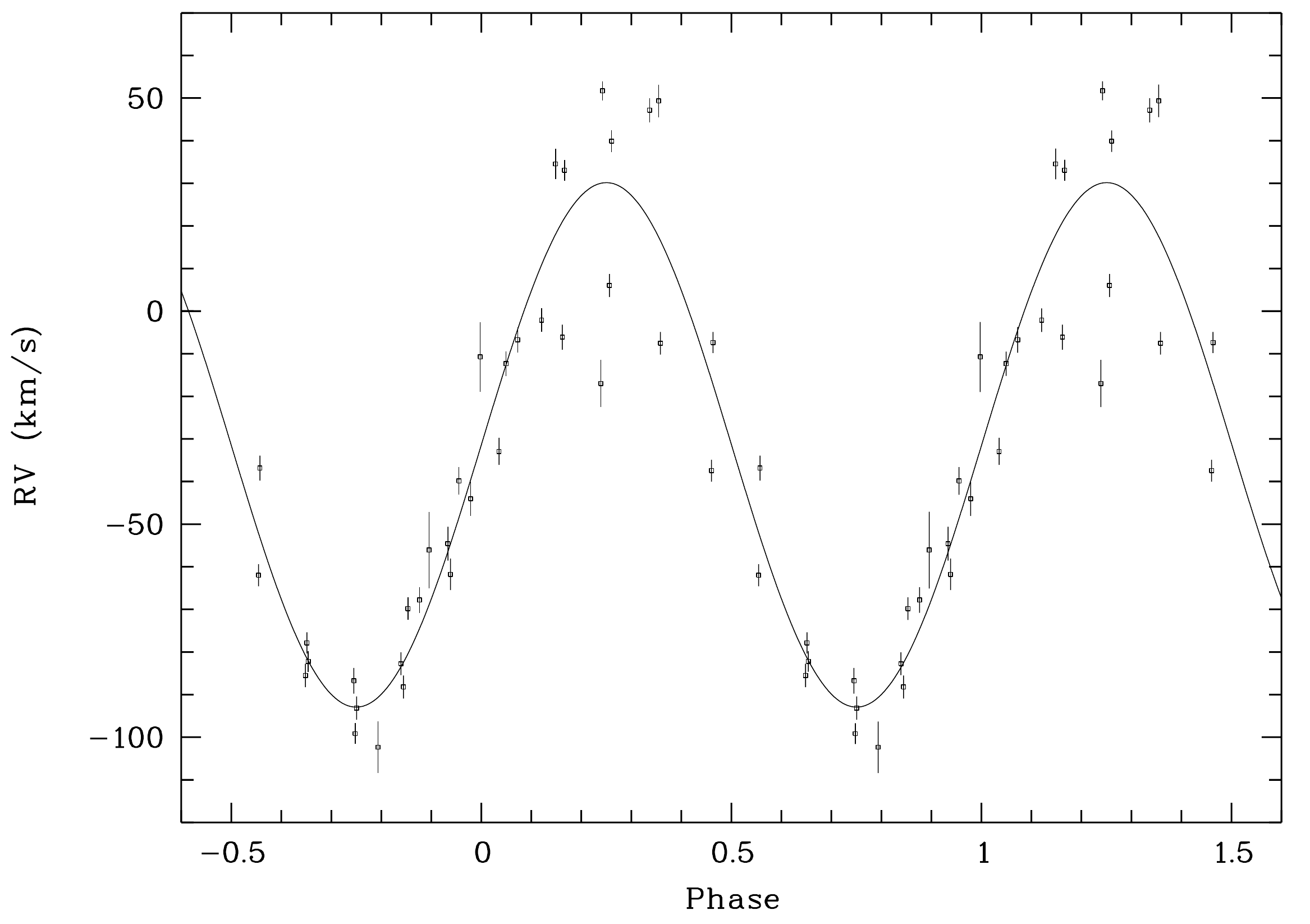}
\caption{Radial velocities of the \Ha line folded with the orbital period 
$\rm P_{\Omega}$=4.033\,hr. The best sine fit is also shown (line).}
\label{sp_velcor}
\end{figure}
The radial velocities of the \Ha\, line, folded at the above best orbital 
ephemeris, are shown in Fig.~\ref{sp_velcor} and the parameters of the 
corresponding best sine fits for the strongest emission lines are given in Table~\ref{velocity}. 
It should be noted that a significant velocity scatter is present around $\phi \sim$ 0.25, at maximum red velocity. We carefully checked by using different sky lines that the effect is not due to a faulty wavelength calibration. In fact, the difference in velocity appears in the two consecutive cycles obtained in the same night (Oct. 25) when negligible  wavelength shifts were observed.
Inspection of the line profiles shows some distortions but an attempt to fit with two gaussians failed to 
give significant results.
The Balmer lines show a $\sim$ 50-60 \kms\, range in velocity amplitudes 
while the HeII lines display a slightly higher value with no significant phase shift. 
%
\section{Discussion}
Out of the 21 sources identified as CVs in the 3rd INTEGRAL catalogue, \igr\, is so far the unique confirmed  IP that was discovered by its hard (E $\geqslant$ 20\,keV) X-ray emission. 
Apart for the close-by dwarf nova SS Cygni and the three polars (V834 Cen and the two asynchronous V1432Aql and BY Cam), all other sources  are already known IPs with however five additional suspected CVs remaining to be identified from their period determinations {\citep {bird07, barlow06, stephen06}}. 
Besides its INTEGRAL high energy emission,  \igr\, has only one known low energy counterpart, the  weak (0.1-2.4 keV) ROSAT source, 1RXS J002258.3+614111 \citep{voges99}.  From a search in the X-ray databases,  a flux is also reported from the RXTE ALL Sky Monitor instrument, in the time interval from February 1996 to March 2007 \citep{levine96}.
We extracted and analysed the (2-10 keV) light curve obtained at a 90 s resolution. The source is not significantly detected with a mean countrate of (0.050$\pm$1.65) c/s and a power spectrum of the light curve does not show any significant peak.  \\
Without any knowledge of the absorption, no attempt was made to fit jointly the ROSAT and INTEGRAL data. From the reported ROSAT hardness ratios, HR1=1.00$\pm$0.09 and HR2=0.51$\pm$0.16 \citep{voges99}, and using the HR-N$_{\rm H}$ distributions for CVs  \citep[see][ Fig. 4 and 5]{motch98},
a very rough estimate of the absorption can be obtained as 
 N$_{\rm H}$ $\sim 2 \times 10^{21}$ $\rm cm^{-2}$, significantly less than the total line-of-sight interstellar value  (N$_{\rm H}$ $\sim 7.3 \times 10^{21}$ $\rm cm^{-2}$).
With this N$_{\rm H}$ value, the extrapolation of the best INTEGRAL fit yields a (0.1-2.4 keV) flux of 
$8.3\times10^{-12}$ \ecs (with nearly a factor 2 uncertainty) . 
Using the count conversion given in  \citet{voges99}, the estimated (0.1-2.4 keV) ROSAT flux is significantly lower at $(1.1\pm0.2)\times10^{-12}$\ecs\, which may indicate that a more significant absorption is present.\\
\igr\, is up to now the faintest CV detected by INTEGRAL.  There is no good indication of the distance 
though a useful  lower limit can be set using the infrared magnitudes from the 2MASS survey and the CV donors sequence recently computed by \citet{knigge06}. For an orbital period of 4.03\,hr, the donor is expected to be an M3 star with an absolute magnitude of M$_K$ $\sim$6.2. If attributed solely to the secondary, the observed K magnitude (K=14.77$\pm$0.12)  \citep{skrutskie06} would then yield a minimum distance of $\sim$520 pc.
To compare with the other high energy IPs detected by INTEGRAL, a distance independent colour-colour diagram  was built by comparing the flux ratios in the ROSAT (L=0.1-2.4\,keV), and the low (M=20-40\,keV) and high (H=40-100\,keV) INTEGRAL energy ranges (see Fig.~\ref{int_color}).
Flux ratios are derived from the ROSAT  {\citep {voges99} and INTEGRAL {\citep {bird07} catalogues for uniformity reasons though fluxes derived from the actual spectral fits may slightly differ  for \igr (this analysis), V709 Cas  {\citep {falanga05} and  IGR J15479-4529  {\citep {bonnet07} for which a more detailed analysis exists. The sources shown in Fig.~\ref{int_color} are the confirmed IPs with known spin and orbital periods.
With H-M= ($0.58\pm0.17$)   and M-L= ($7.23\pm2.40$), \igr\, has characteristics comparable to the hardest sources.\\
The high observed temperature ($kT \approx 31$\,keV) is a lower limit to the maximum shock temperature and therefore indicative of a white dwarf (WD) mass higher than $\sim0.75$M$_{\sun}$\, \citep[see][]{aizu73}. 
Assuming that the Balmer  lines trace the WD orbital motion, the radial velocity amplitude gives a system mass function of
f$_m$ = (3.7$\pm$0.7) $\times 10^{-3}$ M$_{\sun}$. Together with the WD minimum mass, this yields an estimate of the system inclination  i $\geqslant (21-27)\degr$  for a companion mass in the range (0.3-0.7)M$_{\sun}$\,  comptatible with the orbit and a maximum inclination of 37$\degr$ for a 1.4 M$_{\sun}$\,  WD mass.
If a distance of $\sim$ 500 pc is assumed, the unabsorbed luminosity will be L$_x$(0.1-100 keV)$\sim 6.7 \times 10^{32}$ (d/500pc)$^2$ \es, corresponding to a rather low accretion rate of  $\dot{M} \sim 0.51 \times 10^{16}$ \gs for a $\sim0.75$M$_{\sun}$ WD. 
\igr\,, with a ratio P$_{\omega}$/P$_{\Omega}$=0.04, is also in a relatively high degree of asynchronism. 
For the above WD mass and accretion rate, the condition for accretion to take place, corresponding to a magnetospheric radius lower than the corotation one, will imply a magnetic moment lower than  $1.0\times10^{32}$ G\,cm$^3$, corresponding to a surface magnetic field of $\sim 2.4\times10^{5}$ G. Unless the white dwarf is significantly less massive than indicated by the shock temperature, \igr\, could therefore be a particular  case of a low accreting IP with a low magnetic field. 
%
   \begin{figure}
   \centering
\includegraphics[width=8.5cm]{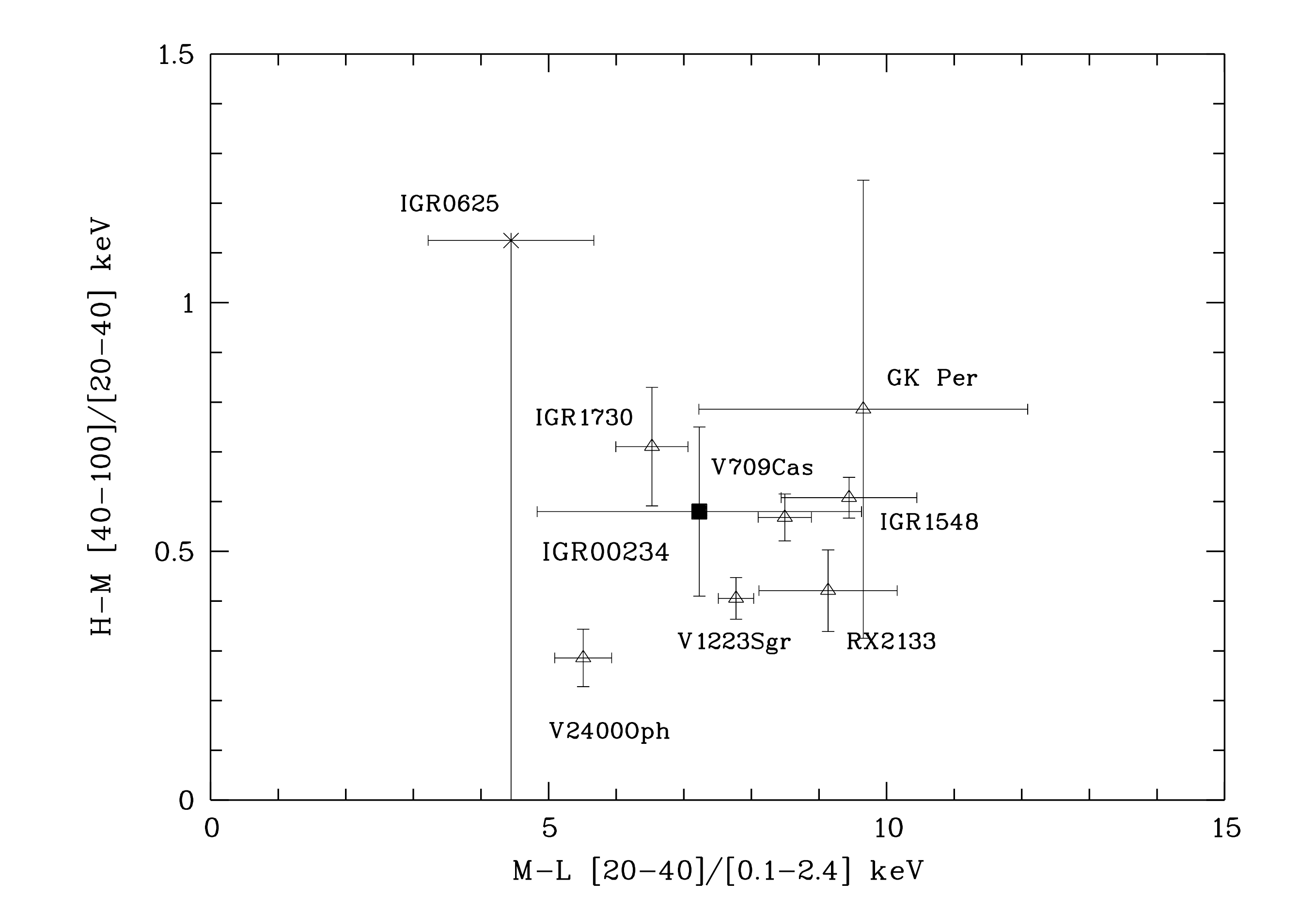}
\caption{High-energy colour-colour diagram of the confirmed IPs detected by INTEGRAL. The energy bands are indicated in the labels.  Sources for which only upper values exist are marked by crosses. \igr (filled square) is among the hardest sources. Note that FO Aqr with M-L=  59.8, due to significant absorption, is  kept out of the scale.}
\label{int_color}
\end{figure}
%
\section{Conclusions}
The analysis of INTEGRAL data has shown that IGR00234+6141 is among the hardest
X-ray emitting CVs with a probable low magnetic field. 
We have identified an orbital variability in the radial velocities of
emission lines at a 4.033\,hr period. 
We have also found a pulsation with a period of 563.5\,s and an optical amplitude of 
0.10\,mag  which is a clear sign of accretion onto a magnetic
white dwarf. This period is identified as the rotational period of the compact object
and should reveal itself as strong X-ray pulsations. The pure sinusoidal modulation and the absence of 
any harmonics is an indication of a one-pole accretion. Due to the faintness of the
source, the INTEGRAL data do not allow a detailed study of the variability 
so that further X-ray observations are clearly needed to confirm these characteristics.
\begin{acknowledgements}
We thank Maria Magri (Obs. Napoli) for her help in collecting photometric data. 
DDM and NM acknowledge the ASI and INAF financial support via grant No. 1/023/05/0.
\end{acknowledgements}

%

\end{document}